\documentclass[prc,preprint]{revtex4}

\usepackage{graphicx}

\begin{document}

\title{\bf The effects of strong magnetic fields on the neutron star structure: lowest
order constrained variational calculations}

\author{Gholam Hossein Bordbar$^{1,2}$\footnote{Corresponding author. E-mail:
bordbar@physics.susc.ac.ir} and Zeinab Rezaei
$^1$}
\affiliation{Department of Physics,
Shiraz University, Shiraz 71454, Iran\\
and \\ Center for Excellence in Astronomy and Astrophysics (CEAA-RIAAM) - Maragha,
P.O. Box 55134-441, Maragha 55177-36698, Iran}
\begin{abstract}We investigate the effects of strong magnetic fields upon
the gross properties of neutron and protoneutron stars. In our
calculations, the neutron star matter was approximated by the pure
neutron matter. Using the lowest order constrained variational
approach at zero and finite temperatures, and employing $AV_{18}$
potential, we present the effects of strong magnetic fields on the
gravitational mass, radius, and gravitational redshift of the
neutron and protoneutron stars. It is found that the equation of
state of neutron star becomes stiffer with increase of the magnetic
field and temperature. This leads to larger values of the maximum
mass and radius for the neutron stars.
\end{abstract}
\maketitle
\section{Introduction}

Compression of magnetic flux inherited from the progenitor star
could form the strong magnetic field in the interior of a neutron
star (Reisenegger ~\cite{Reisenegger}). Using this point of view,
Woltjer has predicted a magnetic field strength of order $10^{15}\
G$ for neutron stars (Woltjer ~\cite{Woltjer}).
In the core of high density inhomogeneous gravitationally bound
neutron stars, the magnetic field strength can be as large as
$10^{20}\ G$ (Ferrer ~\cite{Ferrer}).
In addition, considering the formation of a quark core in the high
density interior of a neutron star, the maximum field reaches up to
about $10^{20}\ G$ (Ferrer ~\cite{Ferrer}; Tatsumi ~\cite{Tatsumi}).
According to the scalar virial theorem which is based on the
Newtonian gravity, the magnetic field strength is allowed up to
$10^{18}\ G$ in the interior of a magnetar (Lai \& Shapiro
~\cite{Lai1}). On the other hand, general relativity predicts the
allowed maximum value of the neutron star magnetic field to be about
$10^{18}\ - 10^{20}\ G$ (Shapiro \& Teukolsky ~\cite{Shapiro}).
By comparing with the observational data, Yuan et al. obtained a
magnetic field strength of order $10^{19}\ G$ for the neutron stars
(Yuan \& Zhang ~\cite{Yuan}).

Strong magnetic field could have an important influence on the
structure of a neutron star. Some authors have studied the effects
of strong magnetic fields on the properties of neutron stars.
Bocquet et al. extended the numerical code for computing the perfect
fluid rotating neutron stars in general relativity to include the
electromagnetic fields and studied the rapidly rotating neutron
stars endowed with magnetic fields (Bocquet et al. ~\cite{Bocquet}).
The results show that for a magnetic field $B \sim 10^{18}\ G$, the
maximum mass increases by $13$ to $29\%$ (depending upon the
equation of state) with respect to the maximum mass of
non-magnetized stars.
Within a relativistic Hartree approach in a simple linear
$\sigma-\omega-\rho$ model, Chakrabarty et al. studied the gross
properties of cold symmetric nuclear matter and nuclear matter in
beta equilibrium under the influence of strong magnetic fields
(Chakrabarty et al. ~\cite{Chakrabarty}). They showed that for
magnetic fields $B_m= 0$, $4.4\times10^{17}$ and $10^{20}\ G$, the
maximum masses are $M_{max} = 3.10M_{\odot}$, $2.99M_{\odot}$ and
$2.91M_{\odot}$, with radii $R_{M_{max}}= 15.02$, $14.95$, $12.25\
km$, respectively.
Based on two nonlinear $\sigma-\omega$ models of nuclear matter,
Yuan et al. considered the properties of neutron stars under the
influence of strong magnetic fields (Yuan \& Zhang
~\cite{Yuan1999}). They found that the equation of state became
softer with increase of the magnetic field. The results show that
for the $ZM$ model, the maximum masses are $M_{max}=1.70M_{\odot}$
and $1.62M_{\odot}$ for $B = 0$, $10^{20}\ G$, with corresponding
radii $R_{M_{max}} = 9.82$ and $8.70\ km$. Furthermore, for the $BB$
model, the maximum masses are $M_{max} = 2.26M_{\odot}$ and
$2.07M_{\odot}$ for $B = 0$, $10^{20}$, with radii
$R_{M_{max}}=12.07$, $10.09\ km$.
Cardall et al. studied static neutron stars with magnetic fields and
a simple class of electric current distributions consistent with the
stationarity requirement (Cardall et al. ~\cite{Cardall}). It has
been demonstrated that the maximum mass of static neutron stars with
magnetic fields determined by a constant current function is
noticeably larger than that attainable with uniform rotation and no
magnetic field.
Within a relativistic field theory, Mao et al. considered a
neutron-star matter consisting of neutrons, protons and electrons
interacting through the exchange of $\sigma$, $\omega$ and $\rho$
mesons in the presence of a magnetic field which decreases from the
center to the surface of a neutron star (Mao et al. ~\cite{Mao}). It
has been found that the equation of state becomes stiffer by
increasing the magnetic field that led to an increase of $40\%$ on
the neutron star maximum mass.

In our previous studies, we have investigated the properties of
neutron stars and protoneutron stars in the absence of magnetic
field (Bordbar et al. ~\cite{Bordbar5991}; Bordbar \& Hayati
~\cite{Bordbar1555}; Bordbar et al. ~\cite{Bordbar61}; Yazdizadeh \&
Bordbar ~\cite{Yazdizadeh}). Recently, we have calculated the
properties of spin polarized neutron matter in the presence of
strong magnetic fields at zero (Bordbar et al.
~\cite{Bordbar044310}) and finite temperatures (Bordbar \& Rezaei
~\cite{Bordbar2011}) using LOCV technique employing $AV_{18}$
potential. In the present work, the neutron star matter is
approximated by the pure neutron matter to investigate the effects
of strong magnetic fields on the gross properties of neuron stars
and protoneutron stars using the equations of state of neutron
matter in the presence of strong magnetic fields (Bordbar et al.
~\cite{Bordbar044310}; Bordbar \& Rezaei ~\cite{Bordbar2011}).


\section{ Neutron star structure in the presence of strong magnetic fields }

\begin{figure}

\includegraphics{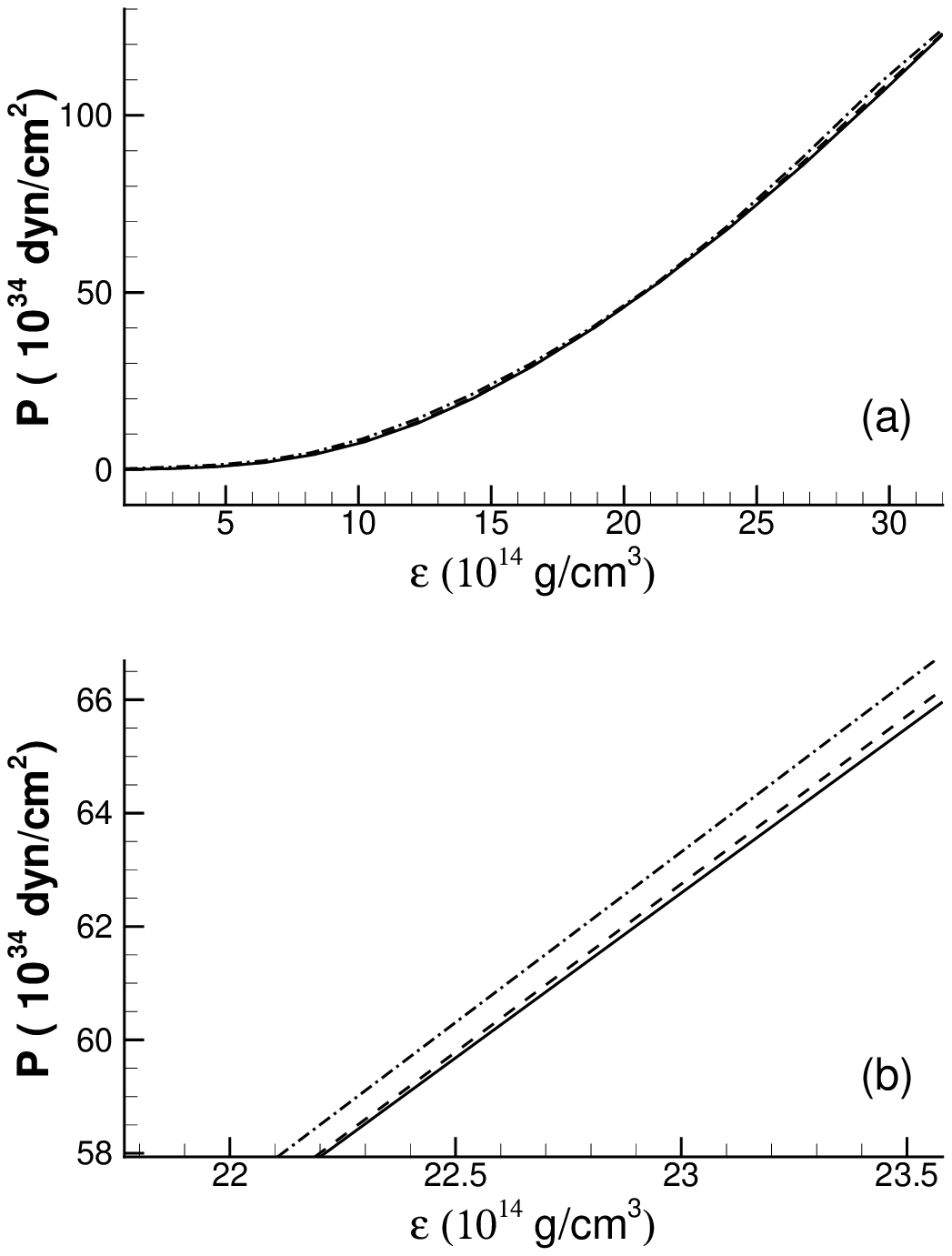}

\caption{\label{fig:11j} (a) Pressure, $P$, versus energy density,
$\varepsilon$, for the cases $B=0\ G$ (solid curve),
$B=5\times10^{18}\ G$ (dashed curve) and $B=10^{19}\ G$ (dashdot
curve) at a fixed value of the temperature, $T=0\ MeV$. (b) Same as
in the top panel but for a different range of density.}

\end{figure}

\begin{figure}

\includegraphics{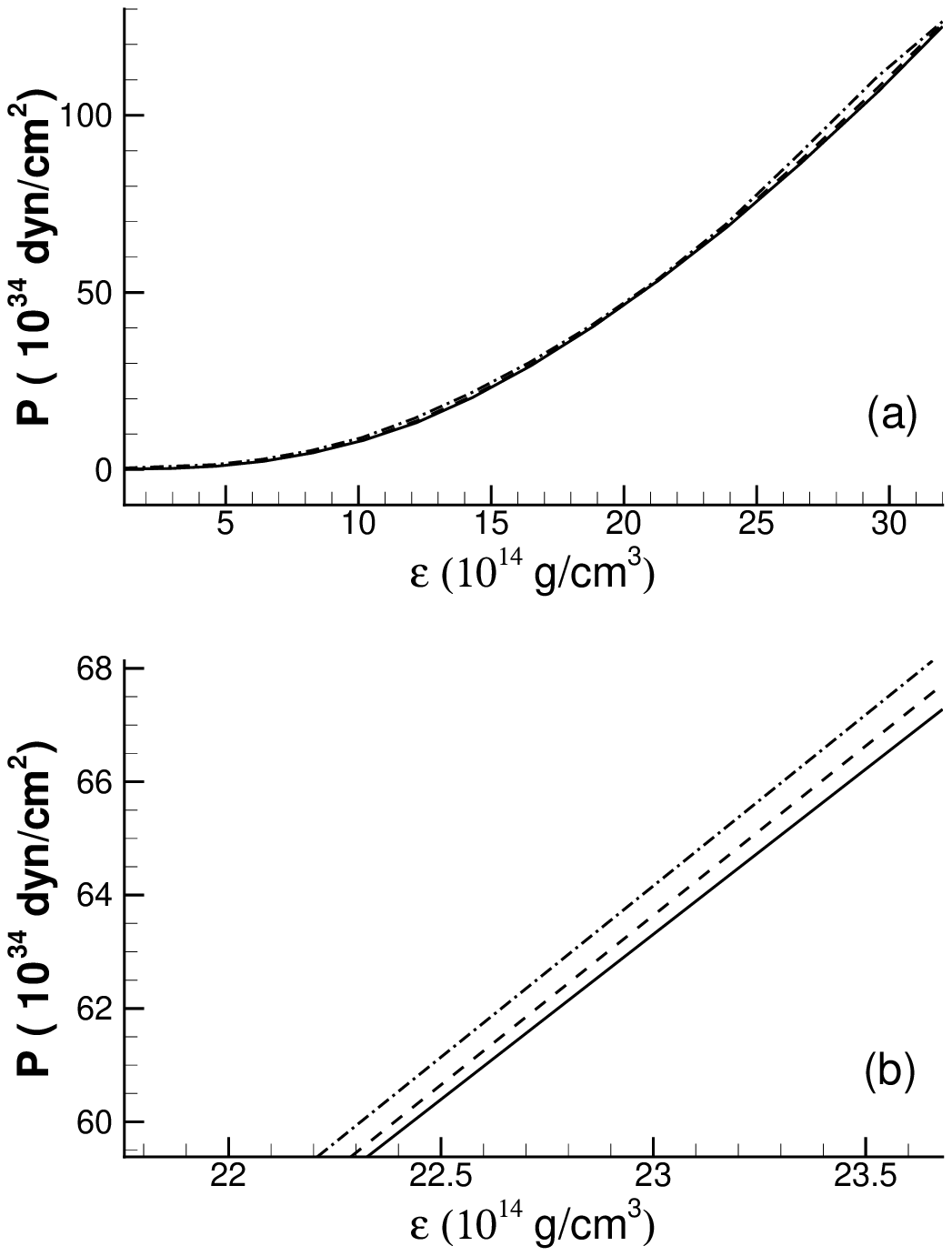}

\caption{\label{fig:12j} (a) Pressure, $P$, versus energy density,
$\varepsilon$, for the cases $B=0\ G$ (solid curve),
$B=5\times10^{18}\ G$ (dashed curve) and $B=10^{19}\ G$ (dashdot
curve) at a fixed value of the temperature, $T=15\ MeV$. (b) Same as
in the top panel but for a different range of density.}

\end{figure}

\begin{figure}

\includegraphics{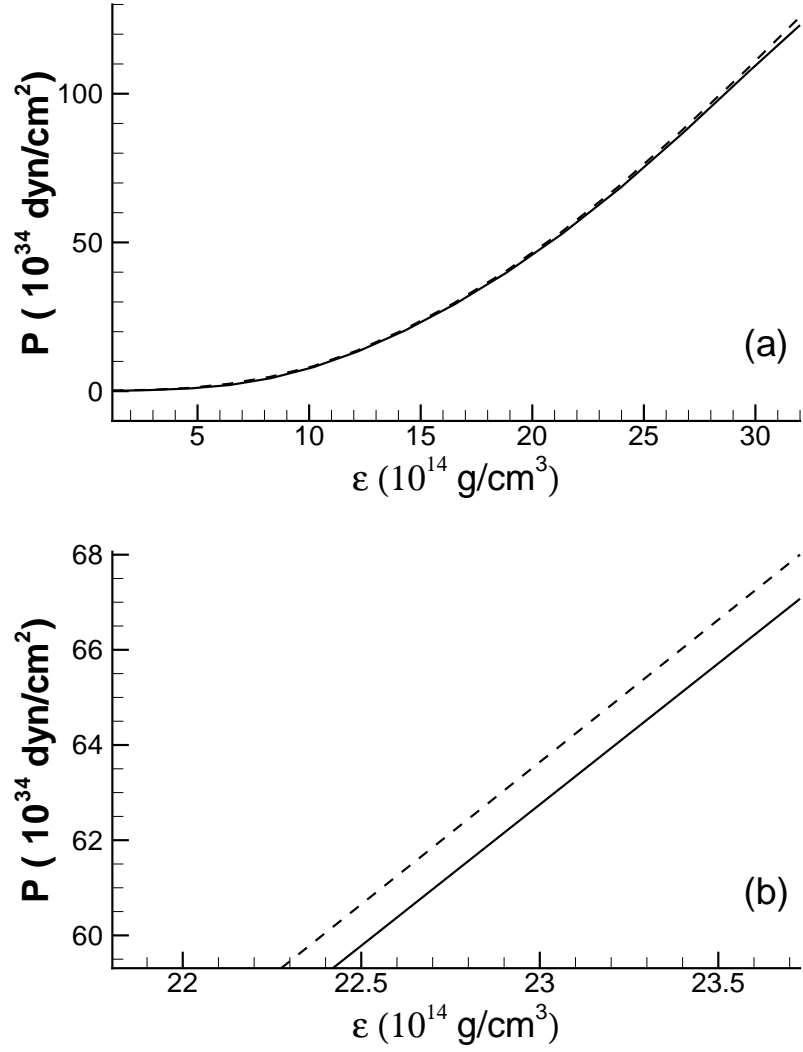}

\caption{\label{fig:13j} (a) Pressure, $P$, versus energy density,
$\varepsilon$, for the cases $T=0\ MeV$ (solid curve) and $T=15\
MeV$ (dashed curve) at a fixed value of the magnetic field,
$B=5\times10^{18}\ G$. (b) Same as in the top panel but for a
different range of density.}

\end{figure}

In the present study, we calculate the neutron star and protoneutron
star structure using the equations of state of cold and hot neutron
matter in the presence of strong magnetic fields (Bordbar et al.
~\cite{Bordbar044310}; Bordbar \& Rezaei ~\cite{Bordbar2011}). In
our study, we employ $AV_{18}$ nuclear potential (Wiringa et al.
~\cite{Wiringa}) and use the lowest order constrained variational
method to calculate the equation of state. For more details, we
refer the reader to (Bordbar et al. ~\cite{Bordbar044310}; Bordbar
\& Rezaei ~\cite{Bordbar2011}). Our results for the equation of
state of neutron matter in the presence of strong magnetic fields
are given in Figs. \ref{fig:11j}-\ref{fig:13j}.
Figs. \ref{fig:11j}(b) and \ref{fig:12j}(b) indicate that for the
cold and hot neutron matter, at each density, the pressure increases
with increase of the magnetic field. This stiffening of the equation
of state is due to the inclusion of neutron anomalous magnetic
moments. In other words, in the presence of high magnetic fields,
the fraction of polarized neutrons at the equilibrium state
increases and therefore the degeneracy pressure grows. This is in
agreement with the results obtained in Refs. (Broderick et al.
~\cite{Broderick}; Yue \& Shen ~\cite{Yue}). We have found that at
low densities, the influence of magnetic field on the pressure is
negligible.
Fig. \ref{fig:13j}(b) shows that at each density, the pressure grows
by increasing the temperature. Consequently, for hot neutron matter,
the equation of state is stiffer compared with the cold one. Fig.
\ref{fig:13j}(a) also shows that the effect of finite temperature on
the equation of state is more significant at high densities.

The equilibrium configurations could be obtained by solving the
general relativistic equations of hydrostatic equilibrium,
Tolman-Oppenheimer-Volkoff (TOV) (Shapiro \& Teukolsky
~\cite{Shapiro}),
\begin{eqnarray}
\frac{dm}{dr} &=& 4\pi r^{2}\varepsilon(r),\nonumber \\
 \frac{dP}{dr}&=&
-\frac{Gm(r)\varepsilon(r)}{r^{2}}\left(1+\frac{P(r)}{\varepsilon(r)c^{2}}
\right)\left(1+\frac{4\pi
r^{3}P(r)}{m(r)c^{2}}\right)\left(1-\frac{2Gm(r)}{c^{2}r}
\right)^{-1},\label{tov1}
\end{eqnarray}
where $\varepsilon(r)$ is the energy density, $G$ is the
gravitational constant, and
\begin{equation}
m(r) =\int_0^r 4\pi r'^2\varepsilon (r')dr'\label{tov2}
\end{equation}
gives the gravitational mass inside a radius $r$. By selecting a
central energy density $\varepsilon_{c}$, under the boundary
conditions $P(0)=P_{c}$, $m(0)=0$, we integrate the TOV equations
outwards to a radius $r=R$, at which $P$ vanishes. This yields the
radius $R$ and mass $M=m(R)$ (Shapiro \& Teukolsky ~\cite{Shapiro}).
Gravitational redshift, the criterion for the star compactness, is
given by
\begin{eqnarray}
Z=[1-2(\frac{GM}{c^2R})]^{-1/2}-1,
\end{eqnarray}
where R is the radius of the neutron star. In our calculations of
neutron star structure, for densities greater than $0.05\ fm^{-3}$,
we use the equations of state presented in Figs.
\ref{fig:11j}-\ref{fig:13j}. However, at lower densities, because
the magnetic field and finite temperature have insignificant effects
on EoS, we employ the equation of state calculated by Baym et al.
(Baym et al. ~\cite{Baym}) for all magnetic fields and temperatures.

\begin{figure}

\includegraphics{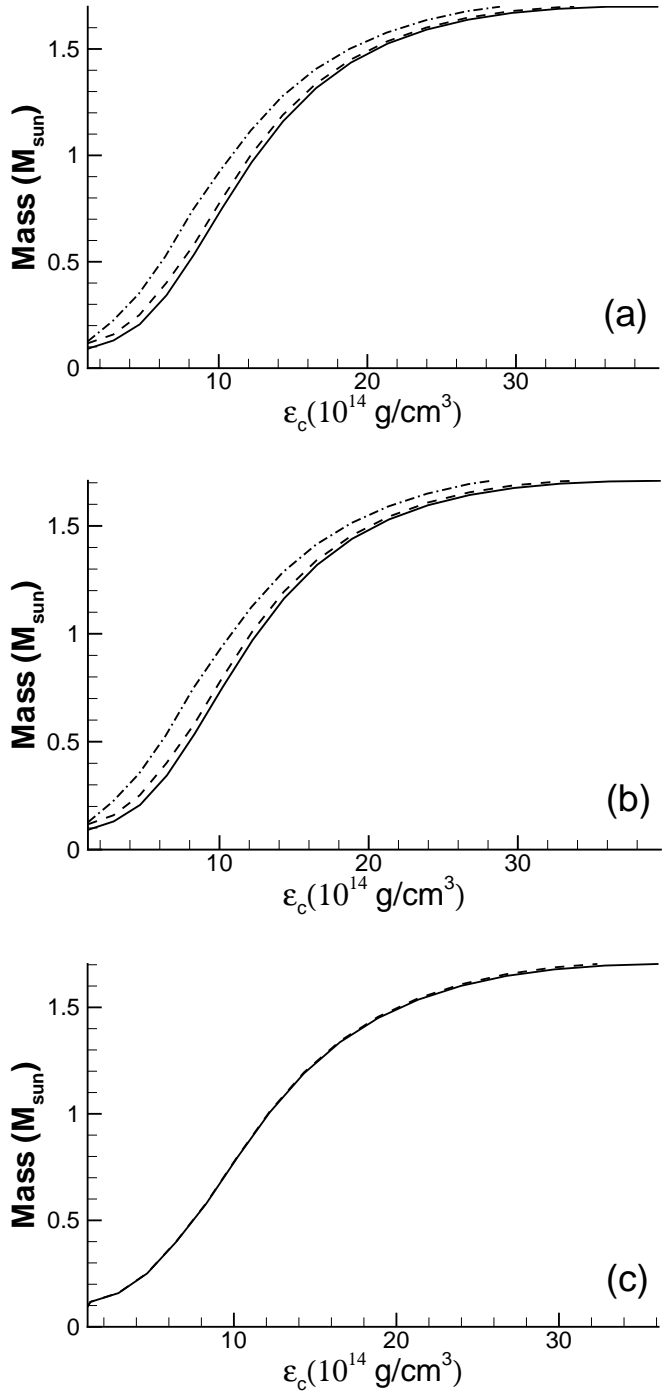}

\caption{\label{fig:2j} (a) Gravitational mass of neutron star (in
units of the solar mass, $M_{\odot}$) versus central energy density,
$\varepsilon_c$, at $T= 0\ MeV$. All curves correspond to those of
Fig. \ref{fig:11j}. (b) Same as (a) but at $T= 15\ MeV$. All curves
correspond to those of Fig. \ref{fig:12j}. (c) Gravitational mass of
neutron star (in units of the solar mass, $M_{\odot}$) versus
central energy density, $\varepsilon_c$, at $B=5\times10^{18}\ G$.
All curves correspond to those of Fig. \ref{fig:13j}.}

\end{figure}

The effects of magnetic fields on the gravitational masses of
neutron stars and protoneutron stars at a temperature about $15\
MeV$ with different central densities are presented in Fig.
\ref{fig:2j}. Obviously, at very low central densities, the
gravitational masses are independent of the equation of state; but
at higher densities, the gravitational mass increases by increasing
both magnetic field and temperature. The limiting value of neutron
star mass (maximum mass) also reaches the larger amount when the
magnetic field and temperature rise.
For a cold neutron star at $B = 10^{19}\ G$, the maximum mass is
about $1.17\%$ larger than the cold non magnetized one. Considering
two stars (a protoneutron star at $T=15\ MeV$ and a cold neutron
star) in the presence of a magnetic field $B = 10^{19}\ G$, the
protoneutron star maximum mass is about $1.16\%$ greater than the
cold neutron star. Besides, for a protoneutron star at $T = 15\ MeV$
and $B = 10^{19}\ G$, the maximum mass increases about $2.36\%$
compared to a cold non magnetized one. These results are due to the
stiffening of the equation of state (Figs. \ref{fig:11j}-\ref{fig:13j}).

\begin{figure}

\includegraphics{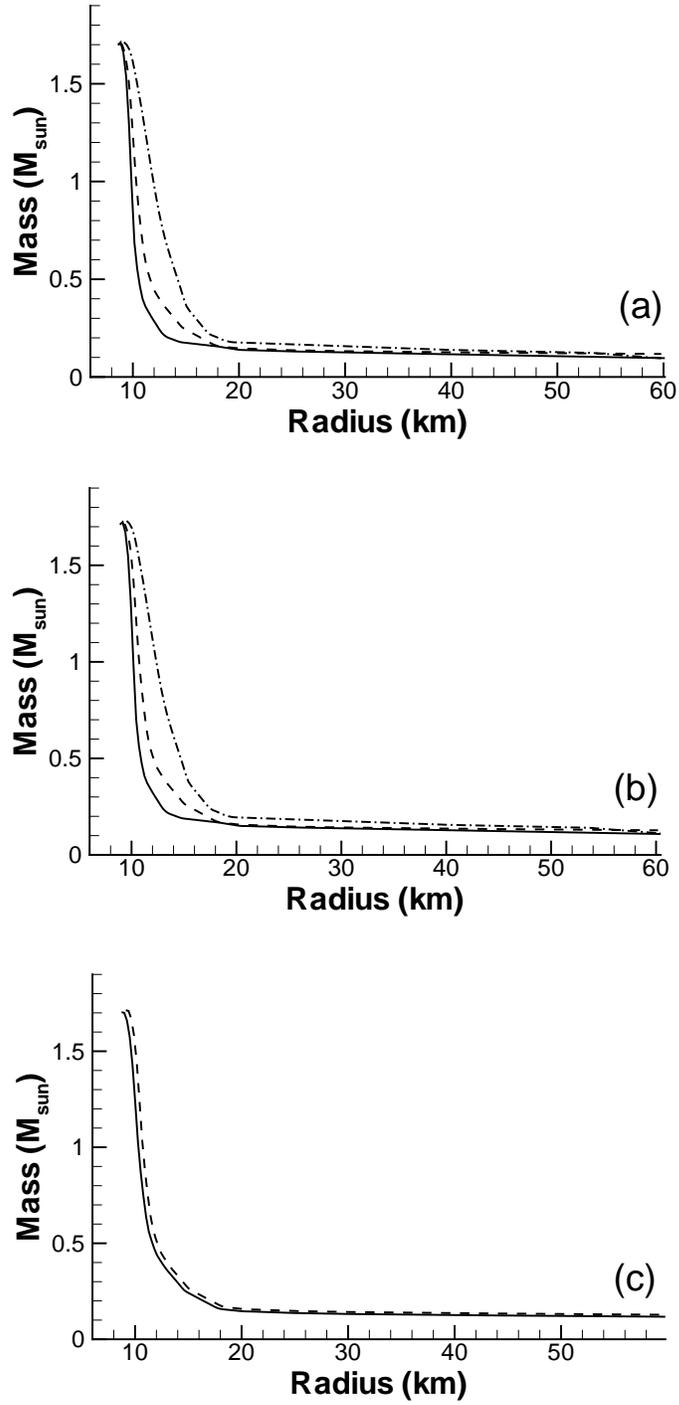}

\caption{\label{fig:3j} (a) Mass-radius relation at $T= 0\ MeV$. All
curves correspond to those of Fig. \ref{fig:11j}. (b) Same as (a)
but at $T= 15\ MeV$. All curves correspond to those of Fig.
\ref{fig:12j}. (c) Mass-radius relation at $B=5\times10^{18}\ G$.
All curves correspond to those of Fig. \ref{fig:13j}.}
\end{figure}

Fig. \ref{fig:3j} presents the gravitational mass versus radius (M-R
relation) for different magnetic fields at zero and finite
temperatures. For all magnetic fields and temperatures, the neutron
star mass decreases by increasing the radius.
It is clear from Fig. \ref{fig:3j} that for a given radius, the
gravitational mass increases whenever the equation of state becomes
stiffer. We have found that the effect of the equation of state on
the M-R relation is more significant for the neutron stars with
smaller radius.

\begin{figure}

\includegraphics{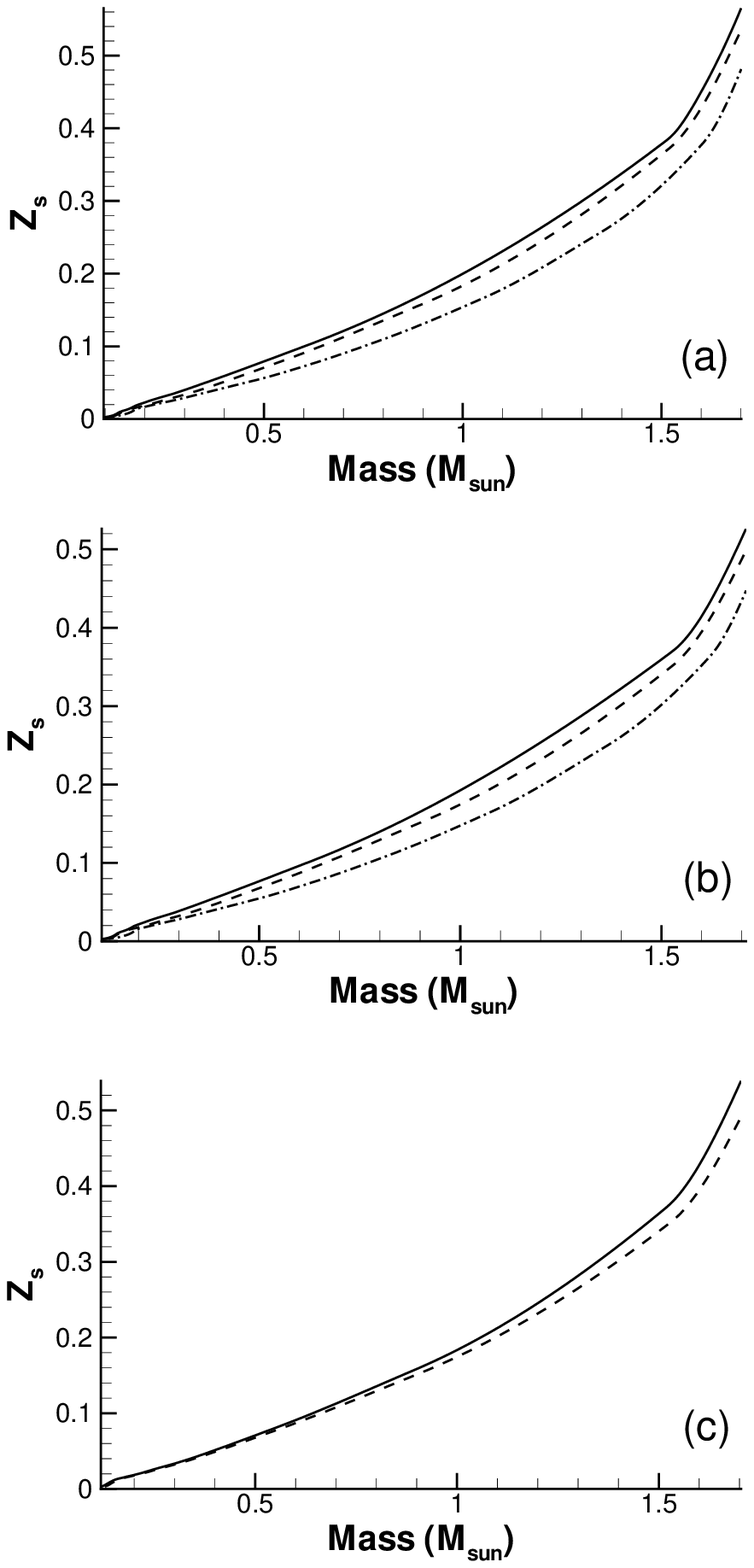}

\caption{\label{fig:4j} Gravitational redshift, $Z_s$, vs. total
mass for neutron stars at $T= 0\ MeV$. All curves correspond to
those of Fig. \ref{fig:11j}. (b) Same as (a) but at $T= 15\ MeV$.
All curves correspond to those of Fig. \ref{fig:12j}. (c)
Gravitational redshift, $Z_s$, vs. total mass for neutron stars at
$B=5\times10^{18}\ G$. All curves correspond to those of Fig.
\ref{fig:13j}.}

\end{figure}

Fig. \ref{fig:4j} shows the gravitational redshift versus the
gravitational mass of the neutron star for different magnetic fields
at zero and finite temperatures.
Clearly, the stiffness of the equation of state reduces the
gravitational redshift. Fig. \ref{fig:4j} also indicates that the maximum
redshift (redshift corresponding to the maximum mass) decreases with
the increase of maximum mass.
For a cold and a hot neutron star at $T = 15\ MeV$ with $B =
10^{19}\ G$, the values of maximum redshift are $z_s^{max}=0.49$, and
$z_s^{max}=0.47$, respectively. In addition, we have found that in
the case of a cold neutron star, for magnetic fields $B = 0$,
$5\times10^{18}$, and $10^{19}\ G$, the values of $z_s^{max}$ are
$0.56$, $0.53$, and $0.49$. Therefore, the maximum surface redshift
of our calculations, i.e. $34.18\%$ (for a cold neutron star at $B =
10^{19} G$), is lower than the upper bound on the surface redshift
for subluminal equation of states, i.e. $z^{CL}_s = 0.8509$
(Haensel et al. ~\cite{Haensel}).

Tables 1 and 2 show a summary of our results for the maximum mass
and the corresponding radius predicted for different neutron stars.
We have found that the effects of magnetic fields with magnitude $B
\leq 10^{18}\ G$ are almost negligible. Obviously, for cold neutron
stars as well as protoneutron stars, the maximum mass and the
corresponding radius increase by increasing the magnetic field.
Tables 1 and 2 show that at any magnetic field, the maximum mass and
the corresponding radius of the protoneuton star are larger than the
cold neutron star. Therefore, we conclude that the stiffer equation
of state leads to a neutron star with a larger maximum mass and
radius. According to our results, for a cold neutron star, the
maximum mass can vary between $1.69 M_{\odot}$ and $1.71 M_{\odot}$,
depending on the interior magnetic field, but for a protoneutron
star with $T = 15\ MeV$, this variation is between $1.70 M_{\odot}$
and  $1.73 M_{\odot}$. Therefore, the effect of magnetic field on
the protoneutron star maximum mass is more important than the cold
neutron star. Our results for the neutron star maximum mass are
higher than the observational results from X-ray binaries presented
in Table 3. Moreover, the study of the statistics of $61$ measured
masses of neutron stars in binary pulsar systems gives a mass
average of $M=1.46\pm0.3 M_{\odot}$ (Zhang et al. ~\cite{Zhang}).
Their results indicate that the mass average of the more rapidly
rotating millisecond pulsars (MSPs) is $M=1.57\pm0.35 M_{\odot}$. In
the present work the values of the protoneutron star radius at
higher magnetic fields are near the values obtained using M–R
relationships (Zhang et al. ~\cite{Zhang2}) which shows the neutron
star radius varies in the range of $10-20\ km$. We have also found
that the effect of magnetic field on the radius of the protoneutron
star is less important than the cold neutron star.


\section{Summary and Conclusion}

Different properties of the neutron star and protoneutron star
structure have been investigated using the equation of state of
 neutron matter in the presence of strong magnetic fields.
In our calculations, we have employed the lowest order constrained
variational method and applied $AV_{18}$ potential to find the
equation of state at zero and finite temperature in the presence of
strong magnetic fields. Our results show that the stiffer equation
of state at higher magnetic fields and larger values of temperatures
lead to the higher values for the maximum mass and radius. For the
maximum value of the magnetic field considered in this study, i.e.
$10^{19}\ G$, the maximum masses of a cold neutron star and
protoneutron star at $T = 15\ MeV$ are $1.71M_{\odot}$ and $1.73
M_{\odot}$, respectively. The corresponding radii are also $9.16$
and $9.22\ km$. Our results indicate that the effects of magnetic
field on the maximum mass of the protoneutron stars are more
important than cold neutron stars, while the effects of magnetic
fields are more visible on the radius of cold neutron stars. It has
been shown that the effects of the equation of state on the M-R
relation are more important for neutron stars with smaller radii.
Our calculations also demonstrate that the maximum value of the
gravitational surface redshift decreases by increasing the neutron
star maximum mass.


\section*{Acknowledgements}
This work has been supported financially by Center for Excellence in Astronomy and Astrophysics (CEAA-RIAAM).
We wish also to thank the Shiraz University Research Council.


\newpage

\begin{table}[h]
\begin{center}
\caption{Maximum gravitational mass, $M_{max}$, and the
corresponding radius, $R_{M_{max}}$, obtained for different values
of magnetic field, $B$, at $T = 0\ MeV$.}

\begin{tabular}{|c|c|c|}
\hline $B (G)$  & \multicolumn{1}{c|}{$M_{max}(M_{\odot})$} &
\multicolumn{1}{c|}{$R_{M_{max}}(km)$}\\\hline $0$ & 1.69&8.59
 \\$5\times10^{18}$ & 1.70  & 8.73 \\
$10^{19}$ & 1.71&9.16\\
\hline
\end{tabular}
\end{center}
\label{tab1}
\end{table}

\begin{table}[h]
\begin{center}
\caption{ Same as Table 1
but at $T = 15 MeV$.}
\begin{tabular}{|c|c|c|}
\hline $B (G)$  & \multicolumn{1}{c|}{$M_{max}(M_{\odot})$} &
\multicolumn{1}{c|}{$R_{M_{max}}(km)$}\\\hline $0$ & 1.70 &8.70
 \\$5\times10^{18}$ & 1.71 & 8.83 \\
$10^{19}$ & 1.73&9.22\\
\hline
\end{tabular}
\end{center}
\label{tab2}
\end{table}

\begin{table}[h]
\begin{center}
\caption{Measured masses of neutron stars in X-ray binaries.}
\begin{tabular}{|c|c|c|}
\hline System  & \multicolumn{1}{c|}{$M(M_{\odot})$} &
\multicolumn{1}{c|}{References}\\\hline SMC X-1 & $1.05\pm0.09$ &
(van der Meer et al. ~\cite{Meer})
 \\Cen X-3 & $1.24\pm0.24$ & (van der Meer et al. ~\cite{Meer}) \\
LMC X-4 & $1.31\pm0.14$&(van der Meer et al. ~\cite{Meer})\\V395
CAR/2S 0921C630 & $1.44 \pm 0.10$&(Steeghs et al.
~\cite{Steeghs})\\
\hline
\end{tabular}
\end{center}
\label{tab2}
\end{table}

\newpage
\newpage

\newpage

\newpage
\newpage
\newpage

\begin{thebibliography}{99}

\bibitem[1971]{Baym} Baym, G., Pethick, C.,  \& Sutherland, P. 1971, Astrophys. J. 170, 299.
\bibitem[1995]{Bocquet} Bocquet, M., Bonazzola, S., Gourgoulhon, E., \& Novak, J. 1995, Astron. Astrophys. 301,
757.
\bibitem[2006a]{Bordbar5991} Bordbar, G.H., Bigdeli, M., \& Yazdizadeh, T. 2006, Int. J. Mod. Phys. A
21, 5991.
\bibitem[2006b]{Bordbar1555} Bordbar, G.H., \& Hayati, M. 2006, Int. J. Mod. Phys. A 21,
1555.
\bibitem[2009]{Bordbar61} Bordbar, G.H., Zebarjad, S. M., \& Zahedinia, R. 2009, Int. J. Theor.
Phys. 48, 61.
\bibitem[2011b]{Bordbar044310} Bordbar, G.H., Rezaei, Z., \& Montakhab, A. 2011, Phys. Rev. C
83, 044310.
\bibitem[2012]{Bordbar2011} Bordbar, G.H., \& Rezaei, Z. 2012, submitted for publication.
\bibitem[2000]{Broderick} Broderick, A., Prakash, M., \& Lattimer, J.M. 2000, Astrophys. J.
537, 351.
\bibitem[2001]{Cardall} Cardall, C.Y., Prakash, M., \& Lattimer, J.M. 2001, Astrophys. J.
554, 322.
\bibitem[1997]{Chakrabarty} Chakrabarty, S., Bandyopadhyay, D., \& Pal, S. 1997, Phys. Rev. Lett.
78, 2898.
\bibitem[2010]{Ferrer} Ferrer, E.J., de la Incera, V., Keith, J.P.,
Portillo, I., \& Springsteen, P.L. 2010, Phys. Rev. C 82, 065802.
\bibitem[1999]{Haensel} Haensel, P., Lasota, J.P., \& Zdunik, J.L. 1999, Astron. Astrophys.
344, 151.
\bibitem[1991]{Lai1} Lai, D., \& Shapiro, S. L. 1991, Astrophys. J. 383, 745.
\bibitem[2003]{Mao} Mao, G., Iwamoto, A., \&  Li, Z. 2003, Chin. J. Astron. Astrophys. 3, 359.
\bibitem[2007]{Reisenegger} Reisenegger, A. 2007, Astron. Nachr. 328, 1173.
\bibitem[1983]{Shapiro} Shapiro, S., \& Teukolsky, S. 1983, \emph{Black Holes, White Dwarfs
and Neutron Stars}, (Wiley,New York).
\bibitem[2007]{Steeghs}Steeghs D., \& Jonker P. G. 2007, ApJ, 669L,
85S.
\bibitem[2000]{Tatsumi} Tatsumi, T. 2000, Phys. Lett. B 489, 280.
\bibitem[2005]{Meer} van der Meer A., Kaper L., van Kerkwijk M. H.,
\& van den Heuvel E. P. J. 2005, AIP, 623.
\bibitem[1995]{Wiringa} Wiringa, R. B., Stoks, V. G. J., \& Schiavilla, R. 1995, Phys. Rev. C 51, 38.
\bibitem[1964]{Woltjer} Woltjer, L. 1964, Astrophys. J. 140, 1309.
\bibitem[2011a]{Yazdizadeh} Yazdizadeh, T., \& Bordbar, G. H. 2011, Res. Astron. Asrtophys. 11, 471.
\bibitem[1998]{Yuan} Yuan, Y. F., \& Zhang, J. L. 1998, Astron. Astrophys. 335, 969.
\bibitem[1999]{Yuan1999} Yuan, Y. F., \& Zhang, J. L. 1999, Astrophys. J. 525, 950.
\bibitem[2006]{Yue} Yue, P., \& Shen, H. 2006, Phys. Rev. C 74, 045807.
\bibitem[2011]{Zhang} Zhang C.M., Wang J., \& Zhao Y.H., et al. 2011, A\&A, 527, 83.
\bibitem[2007]{Zhang2} Zhang C.M., Yin H. X., \& Kojima Y., et al.
2007, Mon. Not. R. Astron. Soc. 374, 232.

%
\end{thebibliography}
\end{document}